\documentclass[prl,twocolumn,showpacs,superscriptaddress]{revtex4}
\usepackage[dvips]{graphicx}
\usepackage{amssymb,amsmath}
\usepackage{dcolumn}
\usepackage{bm}
\usepackage{color}

\graphicspath{{images/}}

\begin{document}

\title{Nonlinear Optical Probe of Indirect Excitons}
\author{A. V. Nalitov}
\affiliation{Institut Pascal, PHOTON-N2, Clermont Universit\'{e}, Blaise Pascal
University, CNRS, 24 avenue des Landais, 63177 Aubi\`{e}re Cedex, France.}
\author{M. Vladimirova}
\affiliation{Laboratoire Charles Coulomb, UMR 5221 CNRS/UM2, Universit\'{e} Montpellier
2, Place Eugene Bataillon, 34095 Montpellier Cedex 05, France}
\author{A. V. Kavokin}
\affiliation{School of Physics and Astronomy, University of Southampton, Southampton,
SO17 1BJ, United Kingdom}
\affiliation{Spin Optics Laboratory, St-Petersburg State University, 1, Ulianovskaya,
St-Peterbsurg, 198504, Russia}
\author{L. V. Butov}
\affiliation{Department of Physics, University of California at San Diego, La Jolla,
California 92093-0319, USA}
\author{N. A. Gippius}
\affiliation{Institut Pascal, PHOTON-N2, Clermont Universit\'{e}, Blaise Pascal
University, CNRS, 24 avenue des Landais, 63177 Aubi\`{e}re Cedex, France.}
\affiliation{A. M. Prokhorov General Physics Institute, RAS, Vavilova Street 38, Moscow
119991, Russia}

\begin{abstract}
We propose the application of nonlinear optics for studies of spatially
indirect excitons in coupled quantum wells. We demonstrate, that despite
their vanishing oscillator strength, indirect excitons can strongly
contribute to the photoinduced reflectivity and Kerr rotation. This
phenomenon is governed by the interaction between direct and indirect
excitons. Both dark and bright states of indirect excitons can be probed by
these nonlinear optical techniques.
\end{abstract}

\pacs{78.20.Bh, 78.67.De, 71.35.-y, 78.47.jg, 78.66.Fd}
\maketitle

\section{I. Introduction.}

Studies of spatially indirect excitons (IX) in semiconductors have attracted
considerable research efforts since early 1960's, fueled by the prediction
of the remarkable quantum properties. This resulted in recent demonstration
of quantum coherent effects including spontaneous coherence \cite{Yang2006,
High2012, Alloing2012}, long-range spin currents and associated polarization
textures \cite{High2013, Vishnevsky2013} of indirect excitons. An IX can be
formed by an electron and a hole confined in separate coupled
quantum wells (CQW). Application of the electric field across the CQWs
bends the band structure so that the IX state became the ground state of the
system \cite{Dzyubenko1996, Sivalertporn2012}. The spatial separation of
electrons and holes within IX allows them to achieve long lifetimes, which
may be orders of magnitude longer than the lifetimes of spatially direct
excitons (DX). At the same time, the spatial separation of electrons and
holes strongly reduces the oscillator strength of IXs,with respect to DXs.
This determines the choice of the experimental methods for studies of these
quasiparticles. The most frequently used optical methods are based on the
emission (photoluminescence, PL) spectroscopy. The PL signal scales linearly
with the emission rate in time-resolved experiments and is nearly
independent on the emission rate in cw experiments for the samples with low
nonradiative recombination. A set of the linear optics methods was employed
for studies of IXs, including the imaging spectroscopy \cite%
{Butov2002}, the time-resolved imaging \cite{Hammack2007}, the
polarization-resolved imaging \cite{High2013}, and the first-order coherence
measurements \cite{Yang2006, High2012, Alloing2012}. However, the powerful
methods of nonlinear optics, which have been successfully applied for DXs in quantum wells (QW) \cite{Schmitt-Rink1989}, \cite{Crooker1996} remain
unexplored in the studies of IXs.

A nonlinear optical process, in its broadest definition, is a process in
which the optical properties of the medium depend on the light field itself
\cite{NonlinearOptics}. In the case of optical pumping in semiconductors,
light-induced modifications of the optical properties of the medium can
persist for a long time after the perturbing light is turned off. In this
case, a pump-probe arrangement can be used, with pump and probe interactions
separated in time \cite{IvchenkoBook}. This allows for time-resolved studies
of optical and spin coherence in the medium. In semiconductor QWs resonant optical pumping of DX resonance with circularly polarized
light, and subsequent detection of the pump-induced dispersive response is
widely used to study exciton population and spin dynamics \cite{baumberg1994}%
. Experimentally, either modification of intensity (photoinduced
reflectivity) or the rotation of the polarization plane of the linearly
polarized probe pulse (photoinduced Kerr rotation) upon reflection from the
sample are measured \cite{Malinowski}. These signals are proportional to the
square of the oscillator strength of the excitonic transition and have a
pronounced resonant character \cite{NonlinearOptics}. Thus, because the
oscillator strength of IX is orders of magnitude lower than for DX, it
is impossible to simply transpose the ideas developed for nonlinear
spectroscopy of DX to IX.

In this paper, we show how IXs, despite their
vanishing oscillator strengths, can induce measurable photoinduced
reflectivity and Kerr rotation. Our proposal relies on two peculiar
properties of the CQW structures. The first essential property is
the spin conserving tunneling of electrons between the QWs
\cite{Poggio}. It allows for substantial spin polarization of IX via
optical orientation of DXs. This has been unambiguously demonstrated
by polarization-resolved photoluminescence experiments
\cite{Leonard2009}. Thus, optical pumping of IXs can de realized via
the DX state. The second imporant effect is the spin-dependent
coupling between DX and IX states. This coupling is quite strong in
CQW, where each IX and each DX have either holes or electrons
located in the same QW. This is why the presence of IX population in
the structure alters DX resonance
properties, mainly via spin-dependent exchange interactions \footnote{%
We note that probing an exciton sub-system with a low oscillator strength
using an exciton sub-system with a higher oscillator strength was earlier
explored in the studies of indirect excitons by linear methods: the
optically dark excitons with spin $\pm 2$ \cite{Maialle1993} or with high
momenta beyond the radiative zone \cite{Feldmann1987, Hanamura1988,
Andreani1991} were probed via the energy shift and optical decay rate of
bright indirect excitons (with spin $\pm 1$ and momenta within the radiative
zone), see e.g. \cite{Butov2001, Hammack2007}.}. Therefore, we suggest that
the detection of IX population and spin polarization can also be realized by
exploiting the DX resonance.
The Kerr rotation measured at the DX resonance is a sensitive method to access populations and spin polarizations of both bright and dark IX states. It may be used, in particular, for studies of dark excitons which strongly affect spin properties of excitonic condensates \cite{High2013,Vishnevsky2013,Combescot2009,Shelykh2012}.
We study the effect of DX-IX interaction
quantitatively, calculating the spin-dependent shifts of the DX resonances
in the presence of spin-polarized gas of IXs within Hartree-Fock and
effective mass approximation. A phenomenological model based on non-local
dielectric response model predicts the spectral dependence of photoinduced
Kerr rotation and reflectivity induced by IXs and allows to analyze the
impact of bright and dark IXs and DXs on these spectra. We also apply spin
density matrix formalism to describe the dynamics of pump-probe signal in a
realistic CQW structure.

This paper is organised as follows. In Section II we describe
phenomenologically the spectral dependence of the reflectivity and
Kerr rotation induced by spin-polarized IXs in the vicinity of DX
resonance. Section III is devoted to the explicit calculation DX-IX
and IX-IX interaction energies. In Section IV we formulate the spin
density matrix model of excitons in CQWs. It accounts for optical
generation of DX excitons, tunneling between the QWs and spin
relaxation. We use the basis of $16$ exciton states (both DX and IX
have four possible spin projections on the growth axis) and
calculate typical reflectivity and Kerr rotation signals as a
function of delay between pump and probe pulses. Section V
summarizes and concludes the paper.

\section{II. Phenomenological model.}

In this Section we analyse phenomenologically the effect of the IX
population in the CQW structure on the polarization and intensity of the
linearly polarized weak probe wave, resonant with the DX transition. Let us
consider a CQW structure schematically shown in Figure 1. It consist of two
QWs, separated by a potential barrier of the width $d$ and covered by a
thick barrier layer of the width $l$. A static electric field is applied
along the $z$-axis, perpendicular to the CQW plane. Two optically-active
ground DX levels in this system with total spin $\pm 1$ are denoted as $%
E_{e} $ and $E_{h}$, and the corresponding wavefunctions as $\Psi _{e}$ and $%
\Psi _{h}$, respectively \footnote{%
Dark DX states characterised by total spin $\pm 2$ have zero oscillator
strength and do not contribute to the reflectivity spectra}. Hereafter, the
indices $e$ ($h$) denote the QW where electrons (holes) are driven by the
static gate voltage. In the case of two identical quantum wells, DX levels
split into symmetrical and antisymmetrical states with very small energy difference  
\cite{Dzyubenko1996} that we will ignore here and assume $E_{e}=E_{h}$.


\begin{figure}[t]
\label{structure}
\center{\includegraphics[width=0.8\linewidth]{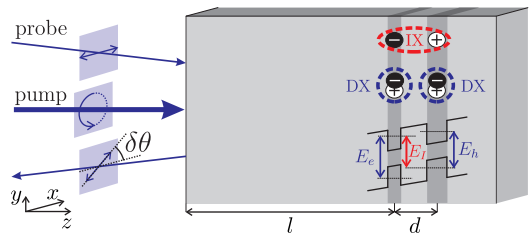} }
\caption{Sketch of pump-probe experiment on CQWs sample. DX and IX
optical transitions are shown. Both pump and probe frequencies are
resonant with one of DX resonance.}
\end{figure}

The electric field of the incident probe beam can be written as $\bm{E}=%
\bm{e}_{x}E_{0}e^{i(kz-\omega t)}$. Here we consider the normally incident
probe beam characterized by the wavevector $k$, propagating along $z$-axis
and linearly polarized along the $x$ axis. The amplitude reflection
coefficient from QW for such wave in the vicinity of one of DX transition
frequencies $\omega _{0}=E_{e(h)}/\hbar $ is related to the exciton
wavefunction $\Psi =\Psi _{e(h)}(\bm{\rho},z_{el},z_{hh})$ by a textbook
formula \cite{Andreani1992,IvchenkoBookNauka} taken in the limit of $kd\ll 1$%
:
\begin{equation}
r_{\mathrm{QW}}(\omega )=\frac{i\Gamma _{0}}{\omega _{0}-\omega -i(\Gamma _{0}+\Gamma
)},  \label{rQW}
\end{equation}%
where the radiative decay rate is given in
\begin{equation}
\Gamma _{0}={\frac{\pi }{2}}k\omega _{LT}a_{B}^{3}S\left[ \int {\Psi (%
\bm{\rho}=0,z,z)dz}\right] ^{2},  \label{Gamma0}
\end{equation}%
$\omega _{LT}$ is the longitudinal-transverse splitting, $a_{B}$ is the bulk
exciton Bohr radius, $S$ is the sample normalization area, and $\Gamma $ is
the exciton non-radiative decay rate, $\rho$ is the in-plane separation of
electron and hole, $z_{el(hh)}$ are electron (hole) $z$ coordinates. %

In the most experimentally relevant case $|r_{\mathrm{QW}}(\omega )|\ll 1$. Taking
into account the interference between waves reflected from the surface of
the cap layer having a refractive index $n$ and those reflected from CQWs,
neglecting re-reflections and reflections from deeper layers of the
structure, the total reflectivity coefficient is
\begin{equation}
r=\frac{1-n}{1+n}+e^{2ikl}\frac{4n}{(n+1)^{2}}r_{\mathrm{QW}}.  \label{structure_refl}
\end{equation}%
%
%
%

\begin{figure}[tbp]
\label{spectra}
\center{\includegraphics[width=0.99\linewidth]{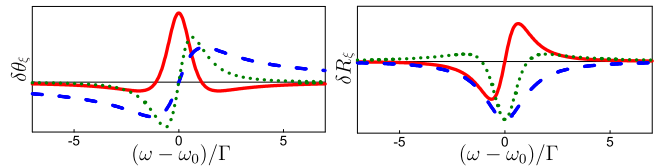}}
\caption{Kerr rotation ($\protect\delta \protect\theta_\xi$) and photoinduced reflectivity ($%
\protect\delta R_\xi$) spectra, calculated from Eq.
(\ref{intens_phi_R})  assuming different nonlinearities: DX energy
shift  ( $\xi=\omega_0$, red solid line), DX transition saturation
 ($\xi=\Gamma_0$, blue dashed line) and DX non-radiative broadening
 ($\xi=\Gamma$, green dotted line).}
\end{figure}

The presence of IXs can affect the DX transition parameters $\omega _{0}$, $%
\Gamma _{0}$ and $\Gamma $ through various mechanisms. Spin-dependent
Coulomb interactions between IXs and DXs leads to the blue shift of DX
levels and, in the case when IXs are polarized, their spin-splitting. IXs
also saturate DX transitions due to the phase space filling effect, since
IXs consist of electrons and holes in the same QWs as DXs. This effect is
again spin-dependent. Finally, IXs may affect the non-radiative decay of DXs
through the scattering processes involving spin-dependent transitions
between DX and IX levels, or by screening the disorder potential. The
renormalisation of exciton resonance frequencies and spin-splittings due to
these interactions is responsible for the modulation of reflectivity and
Kerr rotation spectra \footnote{%
We note that exciton resonance frequency renormalization due to the
interaction with the carriers at other levels was also studied for excitons
in dense electron-hole magnetoplasma \cite{Butov1993} and for bright and
dark indirect excitons \cite{DubinCondmat}.}. Indeed, all the interaction
effects listed above can be accounted for by correcting the reflection
coefficients for two circularly polarized components of the probe pulse, $%
\sigma ^{\pm }$:
\begin{equation}
\delta r^{\pm }={\frac{\partial r}{\partial \omega _{0}}}\delta \omega
_{0}^{\pm }+{\frac{\partial r}{\partial \Gamma _{0}}}\delta \Gamma _{0}^{\pm
}+{\frac{\partial r}{\partial \Gamma }}\delta \Gamma ^{\pm }
\label{correction}
\end{equation}
The electric field of the reflected probe wave can be expressed in terms of
these corrections as
\begin{equation*}
\bm{E}_{\mathrm{r}}=E_{0}{\frac{1}{\sqrt{2}}}\left[ (r+\delta r^{+})\bm{e}%
_{+}+(r+\delta r^{-})\bm{e}_{-}\right]e^{i(kz-\omega t)}.
\end{equation*}%
Here the basis of right and left circularly polarized waves $\bm{e_{\pm}}=(%
\bm{e_x}\pm i\bm{e_y})/\sqrt{2}$ is used. One can see, that in the
most general case, the corrections of the reflection coefficient may
induce (i) circular dichroism, which leads to the build up of
circular polarization, (ii) circular birefringence, which leads to
the rotation of the polarization plane, or Kerr rotation, and (iii)
modification of the probe intensity. In the limit of $|\delta r^{\pm
}|\ll |r_{\mathrm{QW}}|\ll1$ Kerr rotation angle is linear in $(\delta
r^{+}-\delta r^{-})$ \cite{Glazov2012}, while photoinduced
reflectivity is linear in $(\delta r^{+}+\delta r^{-})$:
\begin{equation}
\delta \theta =-\mathrm{Im}\left\{ \frac{\delta r^{+}-\delta r^{-}}{2r}%
\right\}, \ \delta R=|r|^{2}\mathrm{{Re}\left\{ \frac{\delta
r^{+}+\delta r^{-}}{r}\right\} .}  \label{intens_phi}
\end{equation}

Substituting Eqs.(\ref{structure_refl}, \ref{correction}) into Eq.(\ref%
{intens_phi}) we express the three contributions to the photoinduced
Kerr rotation and reflectivity as a function of corresponding
modification of the excitonic characteristic $\xi$, which spans over
the resonant frequency, radiative and non-radiative decay rates $\xi
=\omega _{0},\Gamma _{0},\Gamma $:
\begin{eqnarray}
\delta \theta _{\xi }&=&{\frac{2n}{n^{2}-1}}\mathrm{{Im}\left\{ e^{2ikl}{\frac{%
\partial r_{QW}}{\partial \xi }}\right\} (\delta\xi ^{+}-\delta\xi
^{-})}, \nonumber \\ 
\delta R_{\xi }&=&-{\frac{4n(n-1)}{(n+1)^{3}}}\mathrm{{Re}\left\{
e^{2ikl}{\frac{\partial r_{QW}}{\partial \xi }}\right\} (\delta\xi
^{+}+\delta\xi ^{-})}.  \label{intens_phi_R}
\end{eqnarray}

Figure 2 shows the contributions of different mechanisms to photoinduced
Kerr rotation and reflectivity spectra in the vicinity of the DX resonance,
calculated using Eqs.(\ref{intens_phi_R}), assuming $2kl<<1$ and $n>1$, $%
\delta\omega_0^- = \delta\Gamma_0^- = \delta\Gamma^- = 0$, and
normalized so that $\delta \omega_0^+ / \Gamma = - \delta \Gamma_0^+
/ \Gamma_0 = \delta \Gamma^+ / \Gamma$. This corresponds to: blue
shift of the DX energy $\delta \omega_0$ (red solid line), reduction
of the DX oscillator strength $\delta \Gamma_0$ (blue dashed line)
and enhancement of non-radiative decay of DX, $\delta \Gamma$ (
green dotted line). One can see, that the spectral profiles are
qualitatively different. Thus, measuring photoinduced spectra should
make possible the identification of the underlying nonlinearity.
Note also, that spectral shape depends on the value of the  phase
factor $2kl$ in the Eq. (\ref{intens_phi_R}), so that $\delta\theta$
and $\delta R$ transform as imaginary and real parts of a complex
value between the braces.

The roadmap for the measurement of the pump-probe signal induced by IX and
determination of the underling nonlinearities can be as follows. First of
all, in real time-resolved pump-probe experiments, one operates with short
pulses of light rather than with monochromatic waves. Therefore, to measure
spectral dependence of the nonlinear signal the probe spectral width must be
smaller than the DX linewidth ($\Gamma +\Gamma _{0}$), which is accessible experimentally \cite{Hoffmann2006}. Second, one
should avoid any contribution of DX population to the nonlinear signal. This can be
easily realized by setting the delay between pump and probe pulses
sufficiently long, because DX lifetime does not exceed $500$ ps, while IX
lifetime is at least an order of magnitude longer. Finally, fitting the
measured pump-induced reflectivity and Kerr rotation spectra to Eq.(\ref%
{intens_phi}) assuming different excitonic nonlinearities, it should
be possible to determine the relative importance of different
mechanisms of DX-IX interaction \footnote{%
The Faraday rotation spectroscopy provides the same information as the Kerr
rotation spectroscopy with the only difference in the detection geometry:
the transmitted signal is studied in the Faraday configuration, while in
Kerr configuration the reflected signal is detected.}. In the next Section,
we show that at least one of the discussed mechanisms, namely energy shift
of DX resonance, should produce a measurable nonliner signal in realistic
CQW structures.

\section{III. IX-DX interaction energy}

In this Section we estimate the strength of DX-IX Coulomb interaction in a
typical CQW structure. Following the approach of \cite%
{Ciuti1998,Laikhtman,Shelykh2012} we use the Hartree-Fock and
effective mass approximations to find matrix elements of the Coulomb
interaction Hamiltonian. Our objective is to calculate the energy
shifts of the bright DX levels $E_{e}^{\pm 1}$ and $E_{h}^{\pm 1}$
induced in the first order by the population of both bright
($n_{I}^{\pm 1}$) and dark ($n_{I}^{\pm 2}$) IX states. We also give
estimations for IX-IX interaction energy within the same model, to
compare with the PL line-shifts observed experimentally and with
existing theoretical results.

We operate with a wavefunction of a single 1s-exciton characterized by a
center of mass wave vector $\bm{Q}$ with decoupled translational motion in
the QW plane, electron and hole motion along $z$ axis and relative motion of
electron and hole:
\begin{equation}
\Psi_{\bm{Q}} (\bm{\rho},z_{el},z_{hh})={\frac{1}{\sqrt{S}}}\exp (i\bm{Q}%
\bm{R}_{c.m.})\Psi _{z}(z_{el},z_{hh})\Psi_{\rho}(\bm{\rho}),
\end{equation}%
where $\bm{R}_{c.m.}=(m_{el}\bm{r}_{el}^{||}+m_{hh}\bm{r}%
_{hh}^{||})/(m_{el}+m_{hh})$, $\bm{\rho}=\bm{r}_{el}^{||}-\bm{r}_{hh}^{||}$,
$\bm{r}_{el,hh}^{||}$ are electron and hole radius vector projections on the
CQW plane, $m_{el,hh}$ are electron and hole in-plane effective masses.
Relative motion part of this wavefunction $\Psi _{\rho}(\rho)$ for both DXs
and IXs can be found from the solution of a 2D radial Schroedinger equation
\cite{Gippius1998} or using the variational approach with the trial function
of the form \cite{Shelykh2012}:
\begin{equation}  \label{Psi_rho}
\Psi _{\rho}(\bm{\rho})={\frac{1}{\sqrt{2\pi b(b+r_{0})}}}\exp \left( \frac{-%
\sqrt{\rho ^{2}+r_{0}^{2}}+r_{0}}{2b}\right),
\end{equation}
with $r_0$ and $b$ as variational parameters. The corresponding
mean square of in-plane radius for this exciton wavefunction is given by:
\begin{equation}
p^2 = \frac{2 b }{b + r_0} (r_0^2 + 3 b r_0 + 3 b^2 ). 
\label{radius}
\end{equation}

\begin{figure}[t]
\includegraphics[width=0.9\linewidth]{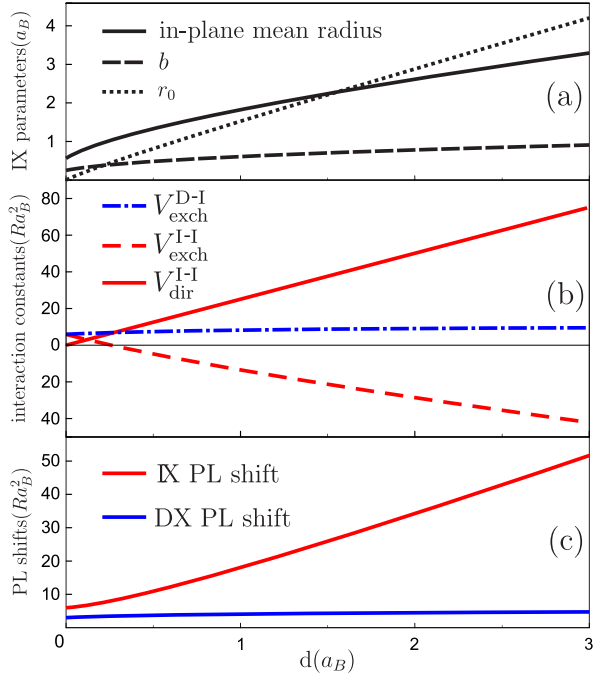}
\caption{ (a) Parameters of the IX wave function obtained via
variational procedure and the resulting in-plane radius. (b)
Interaction constants computed by Monte-Carlo method. DX-IX carrier
exchange constant $V_{exch}^{D-I}$ (blue dot-dashed line), IX-IX
carrier exchange constant $V_{\mathrm{exch}}^{\mathrm{I-I}}$ (red
dashed line) and direct or exciton exchange interaction constant
$V_{\mathrm{dir}}^{\mathrm{I-I}}$ (red solid line). (c) PL shifts of
IX and DX lines calculated from Eqs. \protect \ref{shitsPL}.}
\label{figshifts}
\end{figure}

Figure \ref{figshifts}~(a) shows the $\psi_\rho$  parameters $p$,
$b$ and $r_0$ as a function of  QWs separation $d$, as obtained from
the variational procedure.
In the limiting case of $d=0$, Eq. (\ref{radius}) coincides with the
exact solution for the in-plane radius of the DX wavefunction with $%
r_0=0$ and $b=a_B/4$.
When the separation between QWs increases, the in-plane extension of
the IX wavefunction grows sublinearly.
The wave function in z-direction in the limit of two infinitely thin
QWs reads:
\begin{equation*}
\Psi _{z}^{e(h)}=\delta (z_{el}-Z_{e(h)})\delta (z_{hh}-Z_{e(h)}),
\end{equation*}
for DX state, and  for IX state
\begin{equation*}
\Psi _{z}^{I}=\delta (z_{el}-Z_{e})\delta (z_{hh}-Z_{h}),
\end{equation*}
where $Z_{e(h)}$ are the QW coordinates in the growth direction.

In Born approximation, the renormalization of DX and IX energies
induced by the presence of a thermalized IX gas, is governed by the
Coulomb exciton-exciton interaction operator.
\begin{widetext}
\begin{equation}
\widehat{V}= {\frac{e^{2} }{\epsilon}} \left[ \frac{1}{|\bm{r}_{el}-\bm{r}_{el}^{\prime}|}+ \frac{1}{|\bm{r}_{hh}-\bm{r}_{hh}^{\prime}|}- \frac{1}{|
\bm{r}_{el}-\bm{r}_{hh}^{\prime}|}- \frac{1}{|\bm{r}_{hh}-\bm{r}
_{el}^{\prime }|} \right]
\end{equation}
\end{widetext}

\begin{table}[t]
\begin{tabular}{r|c|c|c|c|}
& $e\downarrow \Uparrow (+1)$ & $e \uparrow \Downarrow (-1)$ & $h \downarrow \Uparrow (+1)$
& $h \uparrow \Downarrow (-1)$
\\ \hline
$I \uparrow \Uparrow (+2) $ & 0 & $V_{\mathrm{exch}}^{\mathrm{D-I}}$ & $V_{\mathrm{%
exch}}^{\mathrm{D-I}}$ & 0
\\ \hline
$I \downarrow \Uparrow (+1)$ & $V_{\mathrm{exch}}^{\mathrm{D-I}}$ & 0 & $V_{%
\mathrm{exch}}^{\mathrm{D-I}}$ & 0
\\ \hline
$I \uparrow \Downarrow (-1)$ & 0 & $V_{\mathrm{exch}}^{\mathrm{D-I}}$ & 0 & $V_{%
\mathrm{exch}}^{\mathrm{D-I}}$
\\ \hline
$I \downarrow \Downarrow (-2)$ & $V_{\mathrm{exch}}^{\mathrm{D-I}}$ & 0 & 0 & $V_{%
\mathrm{exch}}^{\mathrm{D-I}}$
\\ \hline
\end{tabular}%
\caption{Matrix elements contributing to the interaction of four possible IX
spin states with bright DX states.
Indices $e$, $h$ and $I$ refer to DX in electron and hole QWs and IX
states, direction of arrows defines electron ($\uparrow \downarrow$) and
hole ($\Uparrow \Downarrow$) angular momentum projection. }
\end{table}
\begin{table}[t]
\begin{tabular}{r|c|c|}
& $I \downarrow \Uparrow (+1) $ & $I \uparrow \Downarrow (-1)$
\\ \hline
$I \uparrow \Uparrow (+2) $ & $V_{\mathrm{dir}}^{\mathrm{I-I}}$ + $V_{\mathrm{%
exch}}^{\mathrm{I-I}}$ & $V_{\mathrm{dir}}^{\mathrm{I-I}}$ + $V_{\mathrm{exch%
}}^{\mathrm{I-I}}$
\\ \hline
$I \downarrow \Uparrow (+1)$ & 2($V_{\mathrm{dir}}^{\mathrm{I-I}}$ + $%
V_{\mathrm{exch}}^{\mathrm{I-I}}$) & $V_{\mathrm{dir}}^{\mathrm{I-I}}$
\\ \hline
$I \uparrow \Downarrow (-1)$ & $V_{\mathrm{dir}}^{\mathrm{I-I}}$ & 2($V_{%
\mathrm{dir}}^{\mathrm{I-I}}$ + $V_{\mathrm{exch}}^{\mathrm{I-I}}$)
\\ \hline
$I \downarrow \Downarrow (-2)$ & $V_{\mathrm{dir}}^{\mathrm{I-I}}$ + $V_{%
\mathrm{exch}}^{\mathrm{I-I}}$ & $V_{\mathrm{dir}}^{\mathrm{I-I}}$ + $V_{%
\mathrm{exch}}^{\mathrm{I-I}}$
\\ \hline
\end{tabular}%
\caption{Matrix elements contributing to the interaction of four possible IX
spin states with bright IX
states. Direction of arrows defines electron ($\uparrow \downarrow$) and
hole ($\Uparrow \Downarrow$) angular momentum projection. }
\end{table}

Here $\epsilon$ is material permittivity, $e$ is
the electron charge.

The shifts of DXs and IXs energy levels are simply the matrix
elements of the Coulomb exciton-exciton interaction operator over
two-exciton wavefunctions. These wavefunctions are obtained by
antisymmetrization of the product of two single-exciton ground state
($Q=0$) wavefunctions with respect to the permutation of either
electrons or holes.
%
In the general case of two excitons with spin projections on $z$-axis $S$, $%
S^{\prime },$ corresponding to electron spin projections $s_{e}$, $%
s_{e}^{\prime }$ and heavy hole angular momentum projections $j_{h}$, $%
j_{h}^{\prime }$ this average has the following form \cite{Ciuti1998}:
\begin{equation}
V_{SS^{\prime }}=V_{\mathrm{dir}}+\delta _{SS^{\prime }}V_{\mathrm{exch}%
}^{X}+\delta _{s_{e}s_{e}^{\prime }}V_{\mathrm{exch}}^{el}+\delta
_{j_{h}j_{h}^{\prime }}V_{\mathrm{exch}}^{hh},  \label{V}
\end{equation}%
were $\delta _{ij}$ is the Kronecker delta operator. The first term $V_{%
\mathrm{dir}}$ is the direct Coulomb term which corresponds to the classical
electrostatic interaction between the two excitons. $V_{\mathrm{exch}}^{X}$
is the term describing the simultaneous exchange of the two identical
electrons and the two identical holes between two excitons. The third term $%
V_{\mathrm{exch}}^{el}$ is the term due to the electron-electron exchange,
while $V_{\mathrm{exch}}^{hh}$ is the analogous contribution arising from
the hole-hole exchange. In the limit of zero transferred momentum $V_{%
\mathrm{dir}}=V_{\mathrm{exch}}^{X}$ and $V_{\mathrm{exch}}^{el}=V_{\mathrm{%
exch}}^{hh}$ \cite{Ciuti1998, Shelykh2012}.

Table I provides a convenient visual representation of all the interaction
terms in Eq. (\ref{V}). Along the vertical axis all IX spin states are
listed. Four columns show the matrix elements responsible for their
interaction with a pair of bright excitons in each QW. Mutual orientation of
electron and hole spin in each exciton state is shown by arrows. Analogous representation of interactions between all four IX spin states and bright IX states are given by Table II.

For IX-DX interaction it can be shown that $V_{\mathrm{dir}}=V_{\mathrm{exch}%
}^{X}=0$ due to the absence of stationary dipole moment for DXs and the
assumed zero overlap of DX and IX wavefunctions. Therefore energy shifts of
DXs due to IXs are goverened by just one carrier exchange matrix element $V_{%
\mathrm{exch}}^{el}=V_{\mathrm{exch}}^{hh}\equiv V_{\mathrm{exch}}^{\mathrm{%
D-I}}$. As can be seen from Table 1, carrier exchange interaction for bright
IXs is only possible with DXs with the same spin in both QWs, while DXs
interacting with dark IXs have different spin projection signs in left and
right QWs.

In the case of IX-IX interaction direct Coulomb term does not vanish in the
Born approximation due to the oriented dipole moments of IXs and energy
shifts of IXs are expressed in terms of  two interaction constants $V_{\mathrm{dir}%
}=V_{\mathrm{exch}}^{X}=V_{\mathrm{dir}}^{\mathrm{I-I}}$ and $V_{\mathrm{exch%
}}^{el}=V_{\mathrm{exch}}^{hh}\equiv V_{\mathrm{exch}}^{\mathrm{I-I}}$.
Direct Coulomb term is spin independent and enters every line in Table 1,
while carrier exchange between two IXs is only possible if either electrons
or holes have the same spin projections. Two IXs with both electrons and
holes having the same spin projections can also exchange them
simultaneously. This gives factor 2 before $V_{\mathrm{dir}}^{\mathrm{I-I}}$
in corresponding cells of Table I.

Using the Table I, one can write the expressions for the DX and IX
energy shifts, induced by the IX population. The bright DXs energy
shifts depend on the population of the IXs with different spin
projections on the growth axis as:
\begin{equation}
\delta E_{e}^{\pm}=V_{\mathrm{exch}}^{\mathrm{D-I}}(n_{I}^{\pm 1}+n_{I}^{\mp
2}),\;\delta E_{h}^{\pm }=V_{\mathrm{exch}}^{\mathrm{D-I}}(n_{I}^{\pm
1}+n_{I}^{\pm 2}).  \label{shifts}
\end{equation}

Energy shifts of bright IXs are related to IXs populations in a similar way (see Table II):
\begin{eqnarray}
\delta E_{I}^{\pm } = (V_{\mathrm{dir}}^{\mathrm{I-I}}+V_{\mathrm{exch}}^{%
\mathrm{I-I}})(n_{I}^{+2}+n_{I}^{-2})+ \notag \\
+ 2(V_{\mathrm{dir}}^{\mathrm{I-I}}+V_{
\mathrm{exch}}^{\mathrm{I-I}})n_{I}^{\pm 1}+V_{\mathrm{dir}}^{\mathrm{I-I}%
}n_{I}^{\mp 1}  \label{shiftsI}
\end{eqnarray}

Listed interaction constants are the matrix elements of interaction operator
(\ref{V}):
\begin{widetext}
\begin{eqnarray}
V_{\rm{exch}}^{\rm{D-I}}=\int d^3 \bm{r}_{el} d^3 \bm{r}_{hh} d^3
\bm{r}_{el}^\prime d^3 \bm{r}_{hh}^\prime \Psi_0^{\rm{D}}(\bm{r}_{el},\bm{r}_{hh})
\Psi_0^{\rm{I}}(\bm{r}_{el}^\prime,\bm{r}_h^\prime) \widehat{V} (
\bm{r}_{el}, \bm{r}_{hh}, \bm{r}_{el}^\prime, \bm{r}_{hh}^\prime)
\Psi_0^{\rm{D}}(\bm{r}_{el}^\prime,\bm{r}_{hh})
\Psi_0^{\rm{I}}(\bm{r}_{el},\bm{r}_{hh}^\prime), \notag \\
V_{\rm{dir}}^{\rm{I-I}}=\int d^3 \bm{r}_{el} d^3 \bm{r}_{hh} d^3
\bm{r}_{el}^\prime d^3 \bm{r}_{hh}^\prime \Psi_0^{\rm{I}}(\bm{r}_{el},\bm{r}_{hh})
\Psi_0^{\rm{I}}(\bm{r}_{el}^\prime,\bm{r}_h^\prime) \widehat{V} (
\bm{r}_{el}, \bm{r}_{hh}, \bm{r}_{el}^\prime, \bm{r}_{hh}^\prime)
\Psi_0^{\rm{I}}(\bm{r}_{el},\bm{r}_{hh})
\Psi_0^{\rm{I}}(\bm{r}_{el}^\prime ,\bm{r}_{hh}^\prime), \notag \\ 
V_{\rm{exch}}^{\rm{I-I}}=\int d^3 \bm{r}_{el} d^3 \bm{r}_{hh} d^3
\bm{r}_{el}^\prime d^3 \bm{r}_{hh}^\prime \Psi_0^{\rm{I}}(\bm{r}_{el},\bm{r}_{hh})
\Psi_0^{\rm{I}}(\bm{r}_{el}^\prime,\bm{r}_h^\prime) \widehat{V} (
\bm{r}_{el}, \bm{r}_{hh}, \bm{r}_{el}^\prime, \bm{r}_{hh}^\prime)
\Psi_0^{\rm{I}}(\bm{r}_{el}^\prime,\bm{r}_{hh})
\Psi_0^{\rm{I}}(\bm{r}_{el},\bm{r}_{hh}^\prime).
\end{eqnarray}
\end{widetext}

While $V_{\mathrm{dir}}^{\mathrm{I-I}}$ can be found
analytically \cite{Shelykh2012}, $V_{\mathrm{exch}}^{\mathrm{I-I}}$ and $V_{%
\mathrm{exch}}^{\mathrm{D-I}}$ should be calculated numerically
using\textit{ e.g. } Monte-Carlo integration method. The result of
matrix elements calculation by
the Monte Carlo method is shown in Figure \ref{figshifts}~(b) in the units of $%
Ra_{B}^{2}$, where $R$ and $a_{B}$ are bulk exciton Rydberg energy and Bohr
radius, respectively. One can see, that $V_{\mathrm{exch}}^{\mathrm{D-I}}$
almost does not depend on the separation between the QWs (dashed-dotted
line). In contrast, both $V_{\mathrm{exch}}^{\mathrm{I-I}}$ (solid line) and
$V_{\mathrm{dir}}^{\mathrm{I-I}}$ (dashed line) terms increase in absolute
value with increasing distance between QWs, but have different signs. This
result is in a good agreement with the calculations of Refs. \cite%
{Ciuti1998,Shelykh2012}.

To further check the validity of this approach, it is instructive to
calculate using Eqs. (\ref{shifts})-(\ref{shiftsI}) the PL shifts of IX and
DX lines in a typical CQW structure. We consider the situation where most of
IXs are depolarized, so that energy shifts of light emitting states should
be averaged over IX spin projections. This gives:
\begin{eqnarray}
\delta E_{e} &=&\delta E_{h}={\frac{1}{2}}V_{\mathrm{exch}}^{\mathrm{D-I}%
}n_{I} \notag \\
\delta E_{I} &=&\left( {\frac{5}{4}}V_{\mathrm{dir}}^{\mathrm{I-I}}+V_{%
\mathrm{exch}}^{\mathrm{I-I}}\right) n_{I},  \label{shitsPL}
\end{eqnarray}%
where $n_I$ is the total density of IXs. The resulting energy shifts are
shown in Figure \ref{figshifts}~(c) for unitary IX density $n_{\mathrm{I}%
}=a_B^{-2}$. Note that the shifts are obtained assuming negligible
tunnel coupling between QWs and is incorrect in the vicinity of
$d=0$, where transition from weak to strong tunnel coupling occurs.
In the latter case DX PL shift is twice higher. One can see, that
for a given density of IX excitons, the PL shift of IX line exceeds
the DX line shift. This difference dramatically enhances with
increasing separation between QWs. 
For IXs densities $\sim 10^{10}$ $\mathrm{{cm}^{-2}}$
and $d\sim 10$~$\mathrm{nm}$, DX and IX PL shifts are of the
order of $0.1$ $\mathrm{meV} $ and $1$ $\mathrm{meV}$ respectively.
This is consistent with the experimental observations
\cite{Butov2000}.


We can now write down the expressions for both Kerr rotation and
photoinduced reflectivity in the vicinity of each DX resonance. %
Substituting Eq.(\ref{shifts}) into Eq.(\ref{intens_phi_R}) with
$E_{e(h)}$ as $\xi$ we relate these signals to the IX populations:

\begin{eqnarray}  \label{result_asym}
\delta \theta _{e} &=&{\frac{2 n }{n^2-1}} V_{\mathrm{exch}}^{\mathrm{D-I}}%
\mathrm{{Im}\left\{ S_e \right\} %
\left[ n_{I}^{+1}-n_{I}^{-1}+n_{I}^{+2}-n_{I}^{-2}\right] }  \notag \\
\delta \theta _{h} &=&{\frac{2 n }{n^2-1}} V_{\mathrm{exch}}^{\mathrm{D-I}}%
\mathrm{{Im}\left\{ S_h \right\} %
\left[ n_{I}^{+1}-n_{I}^{-1}-n_{I}^{+2}+n_{I}^{-2}\right] }  \notag \\
\delta R_{e} &=&{\frac{4n(1-n)}{(n+1)^3}} V_{\mathrm{exch}}^{\mathrm{D-I}}
\mathrm{{Re}\left\{ S_e \right\} %
 \sum_{s} n_{I}^s  }  \notag \\
\delta R_{h} &=&{\frac{4n(1-n)}{(n+1)^3}} V_{\mathrm{exch}}^{\mathrm{D-I}}%
\mathrm{{Re}\left\{ S_h \right\} %
\sum_{s} n_{I}^s }
\end{eqnarray}
where $s=-2,-1,+1,+2$ numerates IX spin projections and the complex values between the braces are defined by:
\begin{equation} \label{S}
\mathrm{S_{e(h)}(\omega)=e^{2ikl} \frac{\partial r_{QW}}{\partial E_{e(h)}}}
\end{equation}

Numerical application of these formula for IXs densities $\sim 10^{10}$ $%
\mathrm{{cm}^{-2}}$, $d\sim 10$~$\mathrm{nm}$ and $100$ \% spin
polarization of IXs, gives for both photoinduced differential
reflectivity and Kerr rotation angle the values of the order of
$10^{-2}$. This allows predicting a measurable nonlinear signal in
differential reflectivity and Kerr rotation even for weakly spin
polarized IXs. Moreover, measurement of Kerr rotation at both DX
resonances, $E_e$ and $E_h$, allows for determination of spin
polarization degree of dark and bright IX separately: %
\begin{equation}
\mathrm{n_{I}^{+1}-n_{I}^{-1}} \sim \delta \theta_e + \delta \theta_h, \quad
\mathrm{n_{I}^{+2}-n_{I}^{-2}} \sim \delta \theta_e - \delta \theta_h,
\end{equation}
as well as for determination of the total density of IXs:
\begin{equation}
\mathrm{n_{I}^{+1}+n_{I}^{-1}+n_{I}^{+2}+n_{I}^{-2}} \sim \delta R_e \sim \delta R_h
\end{equation}

Note, that the determination of dark IX polarization degree is only possible
for the asymmetric CQW structure, where $|E_{e}-E_{h}|\gg \Gamma $, and two
resonances at direct transitions provide independent signals. In symmetric
CQWs $\delta E_e^{\pm}=\delta E_h^{\pm}$, 
which means that only total IXs density and polarization of bright IXs can
be determined, while polarization of dark IXs remains hidden in this case.
%

In principle, the same effects may be observed at the IX transition
frequency, although its magnitude is proportional to the radiative decay $%
\Gamma_0$ entering Eq.(\ref{rQW}), which decays exponentially for IX
resonance with separation of QWs. Indeed, in the IX resonance case, the
integrand of Eq.(\ref{Gamma0}) describes vanishing tails of the
electron(hole) wavefunction in the hole(electron) QW. Substituting Eq.(\ref%
{shiftsI}) into Eq.(\ref{intens_phi_R}) in a similar manner yields the
following expressions:

\begin{eqnarray} \label{result_indirect}
\delta \theta_{I} &=&{\frac{2 n }{n^2-1}} \left( V_{\mathrm{dir}}^{\mathrm{I-I%
}} +2 V_{\mathrm{exch}}^{\mathrm{I-I}} \right)\mathrm{{Im}\left\{ 
S_I \right\} \left[ n_{I}^{+1}-n_{I}^{-1}\right]}
  \notag \\
\delta R_{I} &=& {\frac{4n(1-n)}{(n+1)^3}} \left( {\frac{3}{2}} V_{\mathrm{dir}}^{\mathrm{%
I-I}} + V_{\mathrm{exch}}^{\mathrm{I-I}} \right) \mathrm{{Re}\left\{
S_I \right\} \left[n_{I}^{+1}+n_{I}^{-1} \right] } + \notag \\
&+&{\frac{4n(1-n)}{(n+1)^3}} \left( V_{\mathrm{dir}}^{\mathrm{I-I%
}} + V_{\mathrm{exch}}^{\mathrm{I-I}} \right) \mathrm{{Re}\left\{ S_I \right\}  \left[ n_{I}^{+2}+n_{I}^{-2} \right]}. 
\end{eqnarray}
where $\mathrm{S_I}$ is defined in the same manner as in Eq.(\ref{S}).

Equations (\ref{result_asym}) and (\ref{result_indirect}) form a closed
non-degenerate system of linear equations on IXs spin state occupancies $%
\mathrm{n_I^s}$, therefore measurements of nonlinear effects on both DX and IX
transition frequencies in asymmetric CQWs allow for resolving all spin
components of IXs system. In particular, bright and dark states populations
can be resolved.

\section{IV. Spin density matrix model}

In this section, we shall specifically describe the time-resolved
optically induced reflectivity and Kerr rotation in CQWs.
The
proposed formalism is similar to the formalism which was
developed and successfully applied to QW microcavities in Ref.
\cite{Masha2006}.
In this type of experiment short and
circularly polarized pump pulse is tuned to direct resonance and
creates DXs with certain spin. Short living DXs either relax into IX
state or recombine, leaving a partially polarized IX system.
Linearly polarized probe pulses, also set to the DX transition
frequency and weak comparing to the pump, act as analyzers of the
current exciton density and polarization.

We use density matrix formalism to model the dynamics of a system containing
both DXs and IXs between pump and probe pulses arrival. The state of the
system can be conveniently described by a 16x16 density matrix with elements
denoted as $\rho _{e,h,e^{\prime },h,^{\prime }}^{s,s^{\prime }}$, where $%
e,h,e^{\prime },h^{\prime }=0,1$ indicate the positions of electron and hole
(0 or 1 for the electron or hole QW, respectively), $s,s^{\prime
}=-2,-1,+1,+2$ is the exciton spin state. This basis is convenient for the
description of the electron and hole tunneling which converts DXs to IXs.

The initial conditions for the density matrix are governed by pump
polarization and frequency. In the considered case of circularly polarized
pumping only diagonal elements of initial density matrix $%
\rho_{e,h,e,h}^{s,s}$, with $s=\pm 1$ for $\sigma^\pm$ pump
helicity, are nonzero. Furthermore, if CQWs are asymmetric, DXs in
both QWs can be pumped independently, so that only one diagonal
component of the initial density matrix is nonzero,
$\rho_{0,0,0,0}^{\pm 1,\pm 1}$ if the pump is tuned to the electron
QW resonance, or $\rho_{1,1,1,1}^{\pm 1,\pm 1}$ if it is set to the
hole QW resonance. In the case of symmetric CQWs both diagonal
components with the same spin are initially nonzero.

The evolution of the density matrix is generally described by a quantum
Liouville equation:
\begin{equation}
i\hbar \frac{\partial \rho }{\partial t}=\left[ \widehat{H},\rho \right]
-i\hbar \widehat{L}\left( \rho \right) ,  \label{Liouville}
\end{equation}
where the Hamiltonian $\widehat{H}$ describes coherent processes,
while Lindblad superoperator $\widehat{L}$ describes all incoherent
processes. We concentrate on the second term of Eq.(\ref{Liouville})
to model relaxational dynamics of the system. We shall account for:
(i) tunneling and energy relaxation of electrons from the hole QW to
the electron QW and of holes vice versa, described by the rates
$\gamma _{e}$ and $\gamma _{h}$ respectively, (ii) radiative decay
of bright DXs in both QWs $\Gamma _{0}^{e}=\Gamma _{0}^{h}=\Gamma
_{0}^D$, the one of bright IXs $\Gamma _{0}^{I}$ and corresponding
nonradiative decays $\Gamma^D$ and $\Gamma^{I}$, (iii) separate spin
flips of electrons and holes, described by spin relaxation rates
$\gamma_{e,s}$ and $\gamma_{h,s}$, and simultaneous electron and
hole spin flips, its rate $\gamma _{X,s}$ is defined by
electron-hole exchange and is only present for DXs since this
mechanism is suppressed for IXs by spatial separation of electron
and hole \cite{Maialle1993, Leonard2009, Vina1999}.
The Lindblad term reads: \begin{widetext}
\begin{align}
L\left( \rho _{e,h,e^{\prime },h,^{\prime }}^{s,s^{\prime }}\right) & =
\left[ \left( \Gamma _{0}^{D}\delta _{e,h}\delta _{e^{\prime },h^{\prime
}}+\Gamma _{0}^{I}\delta _{e,1-h}\delta _{e^{\prime },1-h^{\prime }}\right)
\left( \delta _{s,+1}+\delta _{s,-1}\right) +\Gamma ^{D}\delta _{e,h}\delta
_{e^{\prime },h^{\prime }}+\Gamma ^{I}\delta _{e,1-h}\delta _{e^{\prime
},1-h^{\prime }}\right] \delta _{s,s^{\prime }}\rho _{e,h,e^{\prime
},h,^{\prime }}^{s,s^{\prime }}+  \notag \\
& \gamma _{es}\left[ \delta _{s,+1}\delta _{s^{\prime },+1}\rho
_{e,h,e^{\prime },h^{\prime }}^{+2,+2}+\delta _{s,+2}\delta _{s^{\prime
},+2}\rho _{e,h,e^{\prime },h^{\prime }}^{+1,+1}+\delta _{s,-1}\delta
_{s^{\prime },-1}\rho _{e,h,e^{\prime },h^{\prime }}^{-2,-2}+\delta
_{s,-2}\delta _{s^{\prime },-2}\rho _{e,h,e^{\prime },h^{\prime }}^{-1,-1}%
\right] +  \notag \\
& \gamma _{hs}\left[ \delta _{s,+1}\delta _{s^{\prime },+1}\rho
_{e,h,e^{\prime },h^{\prime }}^{-2,-2}+\delta _{s,-2}\delta _{s^{\prime
},-2}\rho _{e,h,e^{\prime },h^{\prime }}^{+1,+1}+\delta _{s,-1}\delta
_{s^{\prime },-1}\rho _{e,h,e^{\prime },h^{\prime }}^{+2,+2}+\delta
_{s,+2}\delta _{s^{\prime },+2}\rho _{e,h,e^{\prime },h^{\prime }}^{-1,-1}%
\right] +  \notag \\
& \delta _{e,h}\gamma _{Xs}\left[ \delta _{s,+1}\delta _{s^{\prime },+1}\rho
_{e,h,e^{\prime },h^{\prime }}^{-1,-1}+\delta _{s,-1}\delta _{s^{\prime
},-1}\rho _{e,h,e^{\prime },h^{\prime }}^{+1,+1}+\delta _{s,+2}\delta
_{s^{\prime },+2}\rho _{e,h,e^{\prime },h^{\prime }}^{-2,-2}+\delta
_{s,-2}\delta _{s^{\prime },-2}\rho _{e,h,e^{\prime },h^{\prime }}^{+2,+2}%
\right] +  \notag \\
& +\gamma _{e}\delta _{e,0}\delta _{e^{\prime },0}\rho _{1,h,1,h,^{\prime
}}^{s,s^{\prime }}+\gamma _{h}\delta _{h,1}\delta _{h^{\prime },1}\rho
_{e,0,e^{\prime },0}^{s,s^{\prime }}
\end{align}%
\end{widetext}

Only the lowest in energy IX state, for which $e=0$ and $h=1$, is populated
in our model via carrier tunneling and energy relaxation, while the one with
$e=1$ and $h=0$ remains unpopulated and can be safely ignored. Neglecting
nondiagonal elements of the density matrix, we reduce Eq.(\ref{Liouville})
to a linear matrix differential equation on the 12-component vector of DX
and IX spin states populations $n_{e}^{s}=\rho _{0,0,0,0}^{s,s}$, $%
n_{h}^{s}=\rho _{1,1,1,1}^{s,s}$ and $n_{I}^{s}=\rho _{0,1,0,1}^{s,s}$:
\begin{equation}
{\frac{d}{dt}}\left(
\begin{matrix}
n_{e}^{s} \\
n_{h}^{s} \\
n_{I}^{s}%
\end{matrix}%
\right) =\left(
\begin{matrix}
\widehat{L}_{D} & 0 & 0 \\
0 & \widehat{L}_{D} & 0 \\
\gamma _{h}\widehat{I} & \gamma _{e}\widehat{I} & \widehat{L}_{I}%
\end{matrix}%
\right) \left(
\begin{matrix}
n_{e}^{s} \\
n_{h}^{s} \\
n_{I}^{s}%
\end{matrix}%
\right) ,  \label{reducedLiouville}
\end{equation}%
where $\widehat{I}$ is the 4x4 identity matrix, $\widehat{L}_{D}$ and $%
\widehat{L}_{I}$ describe decay and spin relaxation of DXs and IXs,
respectively:
\begin{equation}
\widehat{L}_{D}=\left(
\begin{matrix}
-\Gamma ^{D} & \gamma _{es} & \gamma _{hs} & \gamma _{Xs} \\
\gamma _{es} & -\Gamma _{0}^{D}-\Gamma ^{D} & \gamma _{Xs} & \gamma _{hs} \\
\gamma _{hs} & \gamma _{Xs} & -\Gamma _{0}^{D}-\Gamma ^{D} & \gamma _{es} \\
\gamma _{Xs} & \gamma _{hs} & \gamma _{es} & -\Gamma ^{D}%
\end{matrix}%
\right) ,
\end{equation}%
\begin{equation}
\widehat{L}_{I}=\left(
\begin{matrix}
-\Gamma ^{I} & \gamma _{es} & \gamma _{hs} & 0 \\
\gamma _{es} & -\Gamma _{0}^{I}-\Gamma ^{I} & 0 & \gamma _{hs} \\
\gamma _{hs} & 0 & -\Gamma _{0}^{I}-\Gamma ^{I} & \gamma _{es} \\
0 & \gamma _{hs} & \gamma _{es} & -\Gamma ^{I}%
\end{matrix}%
\right)
\end{equation}%
The solution of Eq.(\ref{reducedLiouville}) can be expressed using matrix
exponent:
\begin{equation}
\left(
\begin{matrix}
n_{e}^{s} & n_{h}^{s} & n_{I}^{s}%
\end{matrix}%
\right) ^{T}=\left. \left(
\begin{matrix}
n_{e}^{s} & n_{h}^{s} & n_{I}^{s}%
\end{matrix}%
\right) ^{T}\right\vert _{t=0}\exp (t\widehat{M}),  \label{matexp}
\end{equation}%
where $\widehat{M}$ is the 12x12 relaxation matrix given explicitly in the
right part of \eqref{reducedLiouville}.

\begin{figure}[t]
\center{\includegraphics[width=0.98\linewidth]{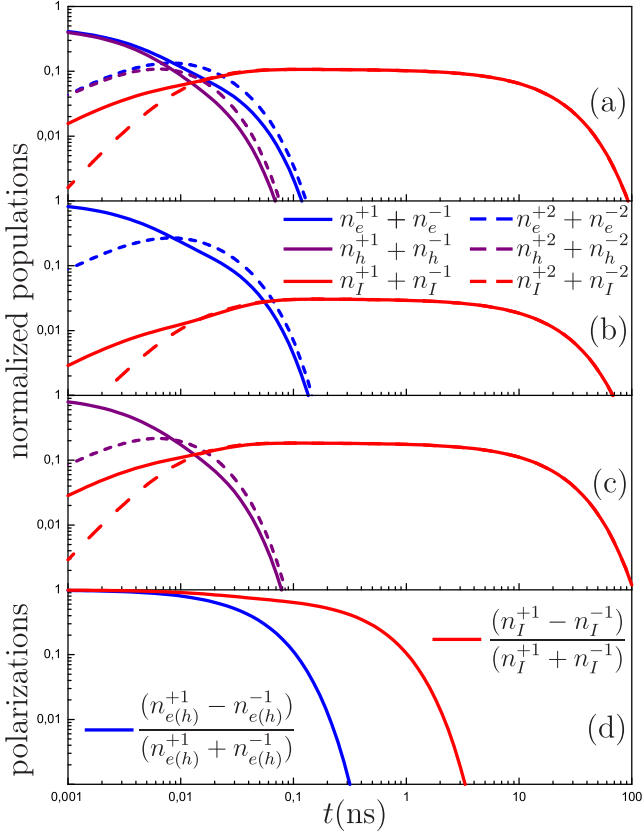} }
\caption{Exciton relaxation dynamics in CQWs after a short circularly
polarized pump pulse: (a) the normalised exciton populations in equally
pumped symmetric CQWs; (b) the same for the pumping of an "electron" QW in
asymmetric CQWs; (c) the same for the pumping of a "hole" QW in asymmetric
CQWs; (d) DX and IX bright exciton polarization degrees. For all panels solid
and dashed curves correspond to bright and dark states, respectively, blue and
purple curves are related to DXs in electron and hole QWs, while red curves describe IXs.
The parameters used in this calculation: $\protect\gamma %
_{h}=(300$ ps$)^{-1}$, $\protect\gamma _{e}=(30$ ps$)^{-1}$, $\Gamma ^{D}\ll
\Gamma _{0}^{D}=(100$ ps$)^{-1}$, $\Gamma ^{I}\ll \Gamma _{0}^{I}=(10$ ns$%
)^{-1}$ \protect\cite{Alexandrou1990}, $\protect\gamma _{es}=(1$ ns$)^{-1}$,
$\protect\gamma _{hs}=(10$ ps$)^{-1}$, $\protect\gamma _{Xs}=(50$ ps$)^{-1}$
\protect\cite{Maialle1994}.}
\label{dynamics}
\end{figure}

Substituting different initial conditions into Eq.(\ref{matexp}) one can
address various experimental scenarios. In the numerical analyses we focus
on the three important cases: (i) DXs in both symmetric QWs
are pumped simultaneously, (ii) CQWs are asymmetric, and we excite
selectively DXs in the electron QW, in which case IXs are formed due to the hole
tunneling, (iii) CQWs are asymmetric, and we excite selectively DXs in the
hole QW, so that IXs are formed due to the electron tunneling. The cases
(ii) and (iii) will be referred to as "electron QW pumping" and "hole QW
pumping", respectively, to emphasize that in the case (ii) we optically
create electrons in the same well where electrons of the lowest energy IXs
are, while in the case (iii) we excite holes in the same well where the
holes of the lowest energy IXs stay.

Figures \ref{dynamics}(a,b,c) show populations of bright and dark DX
and IX states as functions of time in the cases (i,ii,iii),
respectively. In all cases the main features or IX and DX population
dynamics are the same: the IXs population initially increases, while
the DX population shows a fast decay due to the tunneling of
electrons (ii), holes (iii), or both (i), and radiative
recombination of DXs. At longer times, the IX population slowly
decreases. Bright and dark exciton populations quickly equalize due
to the hole spin relaxation. Note that maximum amount of IXs left
after vanishing of DXs depends on the rate of conversion from DXs to
IXs, which is different for listed cases. The conversion due to the
electron tunneling is faster than one due to the hole tunneling
because of the lighter electron than
heavy hole effective mass in the structure growth direction. Figure \ref%
{dynamics}(d) shows the dynamics of polarization degrees of bright DXs and
IXs which is the same for all considered cases. The DX polarization induced
by light quickly decays due to the $\gamma_{X,s}$ relaxation term describing
simultaneous electron and hole spin-flips. IX polarization lives much
longer as for IXs this term is inhibited.\ Interestingly, the fast depolarization of
holes does not lead to the decay of polarization degree of either bright or
dark excitons. Indeed, the transformations from $+1(+2)$ to $-1(-2)$ spin
states or vice versa require both electron and hole spin flips. Fast hole
spin flips, however, do lock bright and dark excitons polarization degrees
to the same absolute values and different signs, which is why we only plot
bright excitons polarization. We note also, that re-polarization of DXs due
to the effective Zeeman splitting induced by polarized IXs is negligibly
small in the chosen range of parameters.

\begin{figure}[h]
\center{\includegraphics[width=0.98\linewidth]{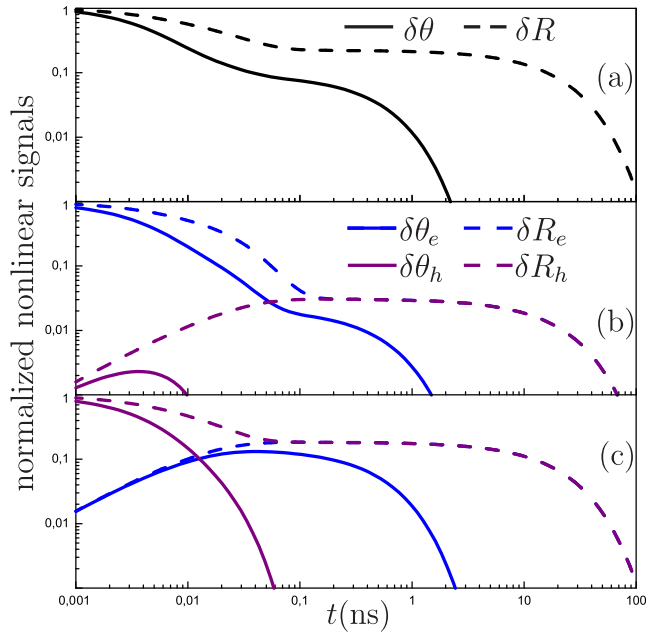} }
\caption{Calculated dynamics of the photoinduced reflectivity (dashed curves) and Kerr
rotation (solid curves) signals in CQWs after a short circularly polarized pump:
(a) symmetric CQWs, both QWs are pumped
(b) asymmetric CQWs, electron QW pumping
(c) asymmetric CQWs, hole QW pumping. Blue and purple lines describe
signals on electron and hole QWs resonant frequencies respectively,
black curves describe folded signals from both symmetric CQWs.
Parameters of calculation are the same as in Figure \protect\ref{dynamics}.}
\label{signals}
\end{figure}

Figure \ref{signals} shows the differential reflectivities and Kerr rotation
angles obtained for asymmetric (a,b) and symmetric (c) CQWs. These
quantities are obtained using Eqs.(\ref{result_asym}) for the IXs
contributions and the following expressions for the DXs contributions
obtained within the same assumptions as in the Section III:
\begin{eqnarray} \label{result_asymDX}
\delta \theta _{e(h)} &=&{\frac{4n}{n^{2}-1}}V_{\mathrm{exch}}^{\mathrm{D-D}}%
\mathrm{{Im}\left\{ S_{e(h)}%
\right\} \left[ n_{e(h)}^{+1}-n_{e(h)}^{-1}\right] }  
\notag \\
\delta R_{e(h)} &=&{\frac{8n(1-n)}{(n+1)^{3}}}V_{\mathrm{exch}}^{\mathrm{D-D}%
}\mathrm{{Re}\left\{ S_{e(h)}%
\right\} \sum_s n_{e(h)}^s},
\end{eqnarray}
where $V_{\mathrm{exch}}^{\mathrm{D-D}}=V_{\mathrm{exch}}^{\mathrm{I-D}%
}(d=0)=V_{\mathrm{exch}}^{\mathrm{I-I}}(d=0)$ are the exchange interaction
constants calculated in \cite{Ciuti1998}. In the present calculation, we
have assumed equality of the interaction constants $V_{\mathrm{exch}}^{%
\mathrm{D-D}}=V_{\mathrm{exch}}^{\mathrm{I-D}}$, which is reasonable
due to the weak dependence of $V_{\mathrm{exch}}^{\mathrm{I-D}}$ on
the distance between the QWs $d$ (see Fig. (\ref{figshifts})). One
can see that the Kerr rotation signal decays faster than the
differential reflectivity signal, in general. This is not surprising
as the Kerr effect is sensitive not only to the population of IXs
but also to their spin polarization, which decays faster than
population. There are two time-scales in the Kerr signal
corresponding to the hole and electron spin relaxation times. In the
case of asymmetric CQWs both reflectivity and Kerr signals are
initially much stronger at the exciton resonance in the pumped QW,
while at the characteristic time-scale of the tunneling transfer the
reflectivity signals from both wells become comparable. The Kerr
signal is always stronger in the electron QW than in the hole QW as
the electron spin relaxation time is much longer than the hole spin
relaxation time. The dynamics of Kerr rotation and differential
reflectivity gives a direct access to DX and IX spin relaxation and
recombination times, but also to the electron and hole spin
relaxation and tunneling times.

\section{\protect\bigskip}

\subsection{V. CONCLUSION.}

We have shown that time-resolved pump-probe experiments  offer new
possibilities for studies of indirect excitons in coupled
quantum well structures.
To circumvent the problem of vanishing oscillator strength of IX
state, we propose to detect the IX population and spin polarization
via modifications of DX resonance properties.
Three different types of nonlinearities due to DX-IX interactions
can be identified.
Namely, DX resonance can shift in energy, change of its oscillator
strength and/or broadening.
We have shown, that relative contribution of these three mechanisms
in measured time-resolved Kerr rotation and reflectivity  can be
identified via spectral profile of the photoinduced signal.

The calculation of DX-IX interaction energy allows to predict
measurable nonlinear signals in CQW structures.
Moreover, in asymmetric CQW structures where two
distinct DX resonances can be addressed, Kerr rotation may provide
information on both bright and dark exciton spin density.

To go further, we built up the spin density matrix formalism
accounting for both DX and IX dynamics, relevant for realistic
pump-probe experiments.
Kerr rotation and reflectivity measured as a function of pump-probe
delay can be described by the model.
Fitting the experimental data to the model should give direct access
to DX and IX decay, electron and hole depolarization and tunneling times.

This work has been supported by the EU FP7 ITN INDEX.



\end{document}